# Structural stability and mechanism of compression of stoichiometric $B_{13}C_2$ up to 68 GPa

Irina Chuvashova[a,b*], Elena Bykova[b], Maxim Bykov[b], Volodymyr Svitlyk[c], Leonid Dubrovinsky[b], Natalia Dubrovinskaia[a]

[a]*Material Physics and Technology at Extreme Conditions, Laboratory of Crystallography, University of Bayreuth, D-95440 Bayreuth, Germany*

[b]*Bayerisches Geoinstitut, University of Bayreuth, D-95440 Bayreuth, Germany*

[c]*European Synchrotron Radiation Facility, BP 220 F-38043 Grenoble Cedex, France*

*\*Correspondence to: irina.chuvashova@gmail.com, natalia.dubrovinskaia@uni-bayreuth.de*

## Abstract

Boron carbide is a ceramic material with unique properties widely used in numerous, including armor, applications. Its mechanical properties, mechanism of compression, and limits of stability are of both scientific and practical value. Here, we report the behavior of the stoichiometric boron carbide $B_{13}C_2$ studied on single crystals up to 68 GPa. As revealed by synchrotron X-ray diffraction, $B_{13}C_2$ maintains its crystal structure and does not undergo phase transitions. Accurate measurements of the unit cell and $B_{12}$ icosahedra volumes as a function of pressure led to conclusion that they reduce similarly upon compression that is typical for covalently bonded solids. A comparison of the compressional behavior of $B_{13}C_2$ with that of α–B, γ–B, and $B_4C$ showed that it is determined by the types of bonding involved in the course of compression. Neither 'molecular-like' nor 'inversed molecular-like' solid behavior upon compression was detected that closes a long-standing scientific dispute.

**Introduction**

Boron carbide $B_4C$ was discovered[1] in 1858, and its crystal structure was first established[2] in 1934. It belongs to icosahedral boron compounds, a family of crystalline solids with crystal structures based on various arrangements of $B_{12}$ icosahedra, which are considered to be some kind of $B_{12}$ "molecules". In such solids chemical bonding has been rationalized in terms of polycenter bonds on the $B_{12}$ *closo*-cluster (boron icosahedron) and two-electron-two-center (*2e2c*) and two-electron-three-center (*2e3c*) bonds between the clusters[3]. According to the Wade-Jemmis rule[4-6] 26 out of 36 valence electrons of a $B_{12}$ cluster are accommodated in 13 molecular-orbital-like bonding orbitals to form the cluster. This leaves 10 electrons for external bonding using 12 equivalent external bonding orbitals, thus creating an electron deficiency in the $B_{12}$ cluster. For a long time, this 'electron deficiency' of the intraicosahedral bonding relative to the intericosahedral one was expected to make the icosahedra more compressible compared to the unit cell, contrary to what is known for typical molecular solids. For this reason the icosahedral boron-rich solids were understood as "inverted-molecular" solids[7]. This gave the origin to the problem of the ratio of the rigidity of the $B_{12}$ *closo*-cluster and the unit cell and the mechanism of compression in icosahedral boron-rich solids that had to be proven experimentally.

The first work, which aimed to shed light into the mechanism of compression of icosahedral boron-rich solids in general and boron carbide in particular, was a high-pressure powder neutron diffraction study of boron carbide ($B_4C$) to 11 GPa conducted by Nelmes and co-authors[8]. By the linear fit of the experimental 'pressure *versus* volume' (P-V) data obtained for both the icosahedron and the unit cell of $B_4C$, the bulk moduli of the icosahedron and the entire structure were determined to be 169(6) GPa and 199(7) GPa, respectively. This led to conclusion that the icosahedron is 23(4) % more compressive than the unit cell of boron carbide, which thus behaved as an "inverted-molecular" solid[8].

An experimental evidence that icosahedral boron materials on compression may not behave as "inverted-molecular" solids came from the single-crystal X-ray diffraction data to 65 GPa obtained for the high-pressure boron allotrope, γ-B[9]. The bulk modulus of the $B_{12}$ icosahedron was determined to be 285(6) GPa, whereas the bulk modulus of the entire γ-B structure was found to be 227(3) GPa[9].

A new insight into the situation in boron carbide, compared to the very first measurements[8], was given due to a compressibility study of 'nearly stoichiometric boron carbide $B_4C$'[10]. Single crystals of $B_4C$ were investigated using synchrotron X-ray diffraction in a diamond anvil cell to 74 GPa. The structure of $B_4C$ was understood as consisting of $B_{11}C$ icosahedra interconnected by the C-B-C chains. The parameters of the equation of state (EoS) of boron carbide were $K_{300}$=243(6) GPa (K´=3.6(2)) or 236(8) at fixed K´=4 ($K_{300}$ is the bulk modulus, and $K´$ is its pressure derivative; the zero pressure unit cell volume was fixed on the ambient pressure experimental value). Dera et al.[10] pointed out that the "icosahedron volume compression did not follow a typical EoS functional behavior" and they did not calculate the bulk modulus of the $B_{11}C$ icosahedron. Nevertheless, they reported that the $B_{11}C$ icosahedron showed a 13% volume reduction, which was smaller than that of the unit cell volume (18 %). The general conclusion was that $B_4C$ behaves as a molecular solid[10] that is in accordance with the results obtained for γ-B[9].

Optical properties and evolution of Raman modes of the same $B_4C$ samples, which were characterized and studied using single-crystal X-ray diffraction in Ref. 10, were also investigated up to 70 GPa by the same research group[11]. Based on their spectroscopy data, the authors[11] reported high-pressure phase transition in boron carbide at 40 GPa, although, according to Dera et al.[10], no signatures of structural phase transitions were observed in $B_4C$ by high-pressure XRD studies up to 70 GPa[10]. Verifying these mutually contradictive reports on boron carbide is not only scientifically important on itself, but it is additionally justified in connection to the commercial value of this unique material. Boron carbide is a very hard and, at the same time, lightweight material for applications in personal security (bullet-proof vests)[3]. It possesses the highest Hugoniot elastic limit of ceramic materials (ca. 17–20 GPa), i.e. the maximum uniaxial dynamic stress that the material can withstand elastically, surpassing all its denser competitors such as silicon carbide and alumina by a factor of 2 (Ref. 12). However, it fails just above the Hugoniot elastic limit and the possible source of failure could be clarified through high-pressure experiments. Thus, establishing the mechanism of compression of boron carbide, clarifying its mechanical properties and limits of its stability under loading, are of both scientific and practical interest.

Here we report high-pressure investigations of stoichiometric boron carbide $B_{13}C_2$ using high-pressure single-crystal X-ray diffraction up to 68 GPa. Single crystals of $B_{13}C_2$, which we study here, were characterized in detail in previous work of our group[13]. As it was established that $B_{13}C_2$ is fully ordered and stoichiometric, and carbon atoms occupy a single position (at the ends of the C-B-C chains)[13], in the present work we could track all changes of the crystal structure, atomic positions, and bond lengths with the high accuracy up to 68 GPa. This enabled us to establish the equations of state of both the $B_{13}C_2$ crystal and the $B_{12}$ icosahedron. As the $B_{12}$ icosahedra in the stoichiometric boron carbide $B_{13}C_2$ do not contain any carbon atoms, contrary to all previously investigated boron carbides[8, 10, 14-24], we could compare the compressional behavior of $B_{12}$ *closo*-clusters in $B_{13}C_2$ and boron allotropes α-B and γ-B and conclude not only regarding the ratio of the rigidity of the $B_{12}$ *closo*-cluster and the unit cell in these materials, but also concerning the mechanism of compression of boron-rich icosahedral solids.

## Results

### *The equations of state of $B_{13}C_2$ and $B_{12}$ icosahedra*

The single-crystal X-ray diffraction data for $B_{13}C_2$ obtained at variable pressure and some experimental details are presented in Table 1. The quality of the data allowed the refinement of both the lattice parameters and atomic coordinates (Supplementary Table S1). The quality of the structural refinement was good up to the highest pressures achieved that gives evidence that in the course of compression the crystals were maintained in quasihydrostatic environment.

In the structure of boron carbide $B_{12}$ icosahedra are located in the corners of the rhombohedral cell, and intericosahedral three-atom C-B-C linear chains are oriented along its body diagonal (Fig. 1a, right). As seen in Figure 1, the structure of $B_{13}C_2$ is very similar to those of α-B and γ-B, which can be described in terms of a cubic closest packing (*ccp*) of spheres, where $B_{12}$ icosahedra play the role of "spheres" (Figure 1a,b). The unit cell parameters of $B_{13}C_2$ in hexagonal setting are $a$ = 5.5962(3) Å, $c$ = 12.0661(7) Å, (space group $R\bar{3}m$), as determined in [13]. Boron atoms in the crystal structure of $B_{13}C_2$ (Figure 1) occupy three crystallographically independent positions ($B_P$, $B_E$, and $B_C$) and the forth position is occupied by carbon atoms in the C-B-C chains[13]. In the present paper we adopt the nomenclature introduced by Ref. 13*:* boron atoms forming the boron icosahedra are labeled as $B_P$ (polar positions) and $B_E$ (equatorial positions), $B_C$ designates the boron atom in the center of the C-B-C chain (Figure 1).

In this study, all observed reflections perfectly match the $B_{13}C_2$ structure up to the highest pressure reached. Our X-ray diffraction data did not reveal any indication of a phase transition. All the unit cell parameters smoothly decrease on compression (Table 1). Figure 2 presents the dependence of the relative unit cell parameters ($a/a_0$ and $c/c_0$) and the relative unit cell ($V/V_0$) volume of $B_{13}C_2$ on pressure up to 68 GPa. As seen (Figure 2), the structure of $B_{13}C_2$ is slightly more compressible along the *c* direction.

The whole experimental volume-pressure data set was fitted using the third-order Birch-Murnaghan (3BM) equation of state that gave the following EoS parameters: $V_0$ = 327.25(4) Å$^3$, $K_{300}$ = 239(7) GPa and $K'$ = 3.2(3) ($V_0$ is the zero pressure unit cell volume, $K_{300}$ is the bulk modulus, and $K'$ is its pressure derivative) (Table 2). The fit with the fixed $K'$ = 4 resulted in the bulk modulus of $K_{300}$ = 222(2) GPa, being lower than that with the free $K'$.

The evolution of the relative volume of the $B_{12}$ icosahedron ($V_{ico}/V_{ico0}$) with pressure is presented in Figure 2. It is very similar to that of the unit cell volume. The whole experimental data set of the icosahedra volumes *versus* pressure (Table 3) was fitted using the 3BM EoS. The EoS parameters were obtained as follows: $V_{0ico}$ = 12.50(3) Å$^3$, $K_{ico}$ = 239(23) GPa, and $K_{ico}'$ =

3.8(8) ($V_{0ico}$ is the zero pressure icosahedron volume, $K_{ico}$ is the bulk modulus of icosahedra, and $K_{ico}'$ is its pressure derivative). The fit with the $K_{ico}' = 4$ (fixed) results in a very close value of $K_{ico} = 234(7)$ GPa. A comparison of the bulk moduli found for the $B_{13}C_2$ crystal and the $B_{12}$ icosahedron (Table 2) shows that the bulk $B_{13}C_2$ and the icosahedron have the similar rigidity.

### *Evolution of the bonds lengths on compression of $B_{13}C_2$*

Recent experimental electron-density study using low-temperature high-resolution single-crystal synchrotron X-ray diffraction data[13] clarified the bonding situation in the stoichiometric boron carbide $B_{13}C_2$. In the present work, we investigated single crystals from the same batch. For consistency, we adopt here the notations of bonds introduced in Ref. 13.

There are seven distinct bonds in the structure of boron carbide (Table 3), which get into three groups[13] (see Figure 1d, bottom, for bonds notations): intra-cluster polycentral bonds ($B_P$–$B_P$, $B_E$–$B_E$, $^1B_P$–$B_E$ and $^2B_P$–$B_E$); inter-cluster bonds ($B_P$–$B_P$), which connect atoms in the polar sites ($B_P$) of the neighboring icosahedra; and bonds involving C-B-C chains (C–$B_E$–and C–$B_C$).

Single-crystal X-ray diffraction data, collected at eleven pressure points in the interval from 4 GPa to 68 GPa, allowed us to follow changes in the length of each of the seven bonds in the structure of boron carbide $B_{13}C_2$ (Table 3). All bonds gradually shorten under compression (Figure 3); their pressure-dependent evolution does not show any anomalies (Figure 3).

### Discussion

As mentioned in the introduction, bulk compressibility of boron carbide compared to compressibility of icosahedra has been a matter of debate. Nelmes et al.[8] reported the crystal structure to be more rigid than the icosahedron cluster, whereas Dera et al.[10] observed an opposite relation.

Our results have shown that the rigidity of the crystal structure of stoichiometric boron carbide $B_{13}C_2$ is similar to that of the $B_{12}$ icosahedron. Within the standard uncertainty, the bulk moduli of the bulk material and the icosahedra we found to be similar: $K_{300} = 239(7)$ GPa ($K' =$ 3.2(3)) *versus* $K_{ico}$ = 239(23) ($K_{ico}'$ = 3.8(8)) (with the fixed $K'$ = 4, $K_{300}$ = 222(2) GPa *versus* $K_{ico}$ = 234(7) GPa). The volume reduction of the unit cell of $B_{13}C_2$ and the volume reduction of the icosahedron in the pressure interval between ambient and 60 GPa were calculated to be also similar (18.7% *versus* 18%, respectively); the difference is less than 1% and within the uncertainty. The bulk modulus obtained by Dera and co-authors[10] for the crystals of $B_4C$ is in a very good agreement with our result: their 3BM EoS parameters are $K_{300}$ = 243(3) GPa, $K'$ = 3.6(6), and $V_0$=330.59(5) Å$^3$ ($V_0$ was fixed at experimental value obtained at ambient conditions); with the fixed $K'$= 4, they got K=236(8) GPa[10]. The volume reduction of the unit cell of $B_4C$ (18%)[10] matches very well to what we obtained for $B_{13}C_2$. But interestingly, the volume reduction of the icosahedron in $B_4C$ in the same pressure interval (between ambient and 60 GPa) in Ref. 10 appeared to be different (13%) (see Table 2).

To clarify a reason of the apparent difference in the ratio of the rigidity of the unit cell and icosahedra found for $B_{13}C_2$ and $B_4C$[10], we first compared the icosahedron and the unit cell volume evolution with pressure in the stoichiometric boron carbide $B_{13}C_2$ and boron allotropes α-B[25] and γ-B[9]. In the structures of each of these materials, icosahedra ($B_{12}$) are built of exclusively boron atoms. Table 2 demonstrates a remarkable observation: in the same pressure range (from ambient up to 60 GPa), the unit cell volume reductions for all these materials are similar (ca. 18% within less than 1% deviation). The $B_4C$, containing $B_{11}C$ icosahedra, is not an exception (18%)[10]. However, the icosahedra volume reductions are all different and reduce in the raw $B_{13}C_2$ (18.1%), γ-B (16.9%), α-B (14.5%), and $B_4C$ (13%). This observation is striking enough and desires an explanation through an insight into the compressional behavior of individual bonds in these solids.

To visualize the difference in the rates of changes of the bonds length and compare boron carbide $B_{13}C_2$ with α-B and γ-B, experimentally obtained data for the relative changes of the bond lengths ($l_P/l_{P0}$) *versus* pressure was linearly fitted for all the bonds ($l_P$ is the length of the bond at pressure P; $l_{P0}$ is the length of this bond at $P_0$ = 4.0 GPa, the first pressure point available in our experiment in the DAC). We plotted calculated "line slopes" against corresponding interatomic distances $l_{P0}$ (Figure 4) at the lowest pressure, similarly to how it was done for characterization of the bond lengths' change under pressure for various boron-rich compounds[26] and α-B[25].

As seen in Figure 4, all the points corresponding to the inter-cluster bonds (C–$B_C$, C–$B_E$, and $B_P$–$B_P$) in $B_{13}C_2$ (black squares) lie on one line, and the rate of their compression depends on the initial length of the bonds (the C–$B_C$ bond between carbon and boron atoms of the C–$B_C$–C chain is the shortest one, see Figure 1a, right panel; Figure 1d, bottom panel; Figure 1b, right panel). Points corresponding to inter-icosahedral bonds in α-B (green circles in Figure 4) drop on the same line: initially shorter $B_P$–$B_P$ bonds contract slower than the initially longer $B_E$–$B_E$ contacts. All points corresponding to the intra-cluster B-B contacts, i.e. those involved in formation of the polycentral bonds of the $B_{12}$ *closo*-cluster, for both $B_{13}C_2$ and α-B (orange squares and red circles, correspondingly, in Figure 4) appear in a very compact area in Figure 4, indicating similar rates of contraction and their similar lengths at ambient conditions. Why then the $B_{12}$ icosahedra in $B_{13}C_2$ and α-B undergo such a dramatically different volume reduction (18.1 *vs* 14.5%) when their crystals are compressed to 60 GPa? Purely geometrical consideration is simply not appropriate. One must take into account types of bonding in each of the solids.

As already mentioned, due to recent experimental electron-density studies of boron allotropes and boron carbide[13, 27, 28], the validity of the Wade-Jemmis model (see introduction) was demonstrated for α–B[27], γ–B[28], and $B_{13}C_2$[13]. In all these solids, the molecular-orbital-type bonding on the icosahedral clusters leaves for *exo*-cluster bonding twelve *sp* hybrid orbitals perpendicular to the surface of the clusters. Thus, in terms of bonding, we can consider $B_{12}$ *closo*-clusters to be similar in these three substances. Concerning the inter-cluster bonds, they were found to be very different in α–B[27], γ–B[28], and $B_{13}C_2$[13]. In α–B and $B_{13}C_2$ the $B_P$–$B_P$ bonds connecting polar $B_P$ atoms (those between neighboring close-packed layers of $B_{12}$ icosahedra) are strong covalent 2e2c bonds (see Figure 1a-c, left and right panels; Figure 1d, top and bottom panels). The $B_E$–$B_E$ contacts in α–B[27] are very flexible, because they are a part of relatively weak 2e3c $B_E$–$B_E$–$B_E$ bonds (shown by brown triangles in Figure 1). Instead of these weak bonds, in the same positions (octahedral voids of the *ccp*) $B_{13}C_2$ possesses three strong 2e2c C–$B_E$ bonds, which are additionally strengthened by the 3e3c bonds of the C–$B_C$–C chains, which impose supplementary negative pressure upon the surroundings, as described in [13]. Now it is clear that the similar contraction of the crystals of α–B and $B_{13}C_2$ (ca. 18%) happens in $B_{13}C_2$ on the expense of the icosahedra, which are set into the very rigid surrounding of strong 2e2c and even stronger 3e3c bonds, but in α–B on the expense of weak 2e3c bonds. This conclusion is justified by the behavior of the unit cell parameters ratio c/a in α–B and $B_{13}C_2$ upon compression: in α–B the c/a ratio increases with pressure, but it decreases in $B_{13}C_2$ (Figure 5). In the both cases, above ca. 40 GPa a clear tendency to the leveling is observed that reflects the more homogeneous compression in the both directions at further pressure increase.

In the light of the consideration made for α–B and $B_{13}C_2$ above, it is not a surprise now that the volume reduction of $B_{12}$ icosahedra in γ–B (16.9%) is intermediate between those in α–B and $B_{13}C_2$ in the same pressure range. As shown in Ref. 28, in γ–B there is a broad diversity of inter-icosahedral bonds and those involving the boron dumbbell (Figure 1a-d, middle panels): three different kinds of strong 2e2c bonds (B3-B3, B2-B5, and B5-B5); for consistency, we follow here the atoms and bonds notations introduced for γ–B in Refs. 9 and 28. Although initially B5-B5 is slightly longer than B3-B3 and B2-B5, it contracts slightly slower than the latter two. The presence of two types (2e3c and 1e2c) of polar-covalent bonds (2e3c B4-B4-B5, involving two boron atoms of one icosahedron and one atom of a dumbbell, and 1e2c B1-B4, involving atoms of neighboring icosahedra at a distance slightly shorter than that between the $B_E$ atoms in α–B) makes the rigidity of the inter-icosahedral space in γ–B to be intermediate between those in α–B and $B_{13}C_2$ (see Figure 1). To compare the pressure evolution of the c/a ratio in α–B and $B_{13}C_2$ with the evolution of the corresponding c´/a´ ratio in the structure of γ–B (c´ is in the direction perpendicular to the close-packed layers and a´ - within the layer), we expressed the c´ and a´ through the a, b and c parameters of the orthorhombic unit cell of γ–B and plotted the c´/a´ ratio *vs* P in Figure 5. As seen, the c´/a´ ratio in γ–B, similarly to what is observed in α–B, first slightly increases, then saturates. Supplementary Figure S1 shows the pressure evolution of the normalized c/a ratios for all of these three solids: the c´/a´(P) curve of γ–B indeed takes an intermediate position between the c/a(P) curves of α–B and $B_{13}C_2$. Thus, the compressional behavior of γ–B is likely controlled by the balance between the contraction of icosahedra and the inter-icosahedral bonds. Figure 4 confirms this interpretation: the lengths of intra-icosahedral contacts in γ–B at ambient pressure are quite similar to those of α–B and $B_{13}C_2$ (see positions of purple triangles in Figure 4), whereas rates of their contraction under pressure vary considerably (see their scatter within the black rectangular outlining the intra-icosahedral bonds in Figure 4). Some of them are as compliant as inter-icosahedral contacts (B1-B4 and B5-B5), whereas others are almost as stiff as 2e2c bonds inter-icosahedral and dumbbell bonds of γ–B itself and 2e2c bonds of boron carbide.

Coming back to the difference in the compressibility of icosahedra in $B_4C$[10] and stoichiometric boron carbide $B_{13}C_2$, one should take into account different chemical composition, i.e. the boron to carbon ratio, of these two carbides. It is established that boron carbide exists as a single-phase material with a wide homogeneity range of carbon content, from ~7 at. % ($B_{14}C$) to ~20 at. % ($B_{4.3}C$), realized by the substitution of boron and carbon atoms for one another within both the icosahedra and intericosahedral chains[16]. Proposed stoichiometries are based on two extreme models, $B_{12}$(CBB) on the boron-rich and $B_{11}$C(CCC) on carbon-rich ends. The location of C atoms in the crystal structure, which is a long-standing debate [10, 14-24], is difficult to clarify even on single-crystal diffraction data because of the similarity in both electronic and nuclear scattering cross-sections for boron and carbon.

For their "nearly stoichiometric boron carbide $B_4C$" Dera et al.[10] suggested the presence of 85 atomic % boron and 15 atomic % carbon in the $B_P$ positions in $B_{11}C$ icosahedra. The presence of carbon likely has to lead to an increase of the rigidity of the icosahedron *closo*-cluster of $B_4C$ due to the higher electronegativity of carbon compared to boron. Indeed, the volume reduction of the $B_{11}C$ icosahedron (13%)[10] appears to be smallest compared to $B_{12}$ icosahedra in boron allotropes α–B (14.5%) and γ–B (16.9%) (see Table 2). The stoichiometric boron carbide $B_{13}C_2$ studied in the present work has been proven to contain carbon only in the C-B-C chains[13], so that its $B_{12}$ icosahedra are much more compliant than $B_{11}C$ (Table 2), moreover their compliance is similar to that of the whole structure. This observation gives additional evidence that the stoichiometric boron carbide $B_{13}C_2$ is a compound with true covalent bonding: $B_{12}$ icosahedra do not play a role of "molecules", their conventional separation is surely convenient for geometric presentation of the structure, but the compressional behavior of the stoichiometric boron carbide is governed by a complex interplay of both intra-cluster and inter-cluster bonds, as well as those involving C-B-C chains.

To conclude, in the present work we have studied the compressional behavior of the stoichiometric boron carbide $B_{13}C_2$ in the pressure interval up to 68 GPa. Our single-crystal synchrotron X-ray diffraction investigations revealed structural stability of the boron carbide in the studied pressure range. A comparison of the unit cell volume reduction with the reduction of the volume of the $B_{12}$ icosahedron upon compression of $B_{13}C_2$ from ambient pressure to 60 GPa revealed their similarity. This confirms that the stoichiometric boron carbide $B_{13}C_2$ is a true covalent compound and does show neither 'molecular-like' nor 'inversed molecular-like' solid behavior upon compression, as previously disputed. Our analysis has shown that, in agreement with the modern understanding of bonding in α–B, γ–B, and $B_{13}C_2$ based on the experimental electron-density studies[13, 27, 28], the compressional behavior of these boron allotropes and boron carbide depends on the types of bonding involved in the course of compression, so that the 'effective compressibility' of $B_{12}$ icosahedra may vary in a broad range, from ca. 14% in α–B to ca. 18% in $B_{13}C_2$, as compared to ca. 18% of compression of the corresponding crystals.

**Methods summary**

*Synthesis of crystals*

Single crystals of $B_{13}C_2$ were synthesized at high pressures (8.5–9 GPa) and high temperatures (1873–2073 K) using the large-volume-press technique. The stoichiometric composition $B_{13}C_2$ was determined by single-crystal X-ray diffraction in agreement with energy-dispersive X-ray (EDX) analysis ($B_{6.51(12)}C$)[13]. The presence of impurities could be excluded.

*Diamond-anvil cell experiments*

The BX90-type diamond anvil cells (DAC)[29] made at Bayerisches Geoinstitut (Bayreuth, Germany) and diamonds with the culet diameters of 250 µm were used in high pressure experiments. Rhenium gaskets were squeezed between the diamonds to make an indentation with the thickness of 30 µm. Then in the center of indentations, round holes of 120 µm in diameter were drilled. The $B_{13}C_2$ crystals were placed into these chambers. Sizes of the crystals were about $10 \times 10 \times 15$ µm³ and $20 \times 15 \times 10$ µm³. Neon was used as a pressure transmitting medium. Ruby balls used for pressure calibration were placed into the pressure chamber.

*Single-crystal X-ray diffraction*

Crystals with size of about $10 \times 10 \times 15$ µm³ were selected for measurements in a DAC at ID27 at the European Synchrotron Radiation Facility (ESRF). Diffraction data were collected at 293 K using the Perkin Elmer XRD1621 flat panel detector. The monochromatic radiation had the wavelength of 0.37380 Å and the crystal-to-detector distance was 383 mm. Pressure in the cell was increased up to 68 GPa with a step of about 6 GPa. 160 frames in the omega scanning range of −40° to +40° were collected (0.5° scanning step size) with an exposure time of 1 s. Integration of the reflection intensities and absorption corrections were performed using CrysAlisPro software[30, 31]. The structure was refined in the anisotropic approximation for all atoms by full matrix least-squares using Jana software[32].

**Additional information**

**Accession codes:** The X-ray crystallographic coordinates for structures reported in this article have been deposited at the Inorganic Crystal Structure Database (ICSD)[33] under deposition number CSD 432649-432659 (compression from 4 to 68 GPa). These data can be obtained free of charge from Fachinformationszentrum Karlsruhe, 76344 Eggenstein-Leopoldshafen, Germany (Fax: +49 7247 808 666; e-mail: crysdata@fiz-karlsruhe.de, http://www.fiz-karlsruhe.de/request_for_deposited_data.html).

**Figure 1. Crystal structures of α-B, γ-B and $B_{13}C_2$.** **(a.)** Rhombohedral cells of α-B and $B_{13}C_2$ compared to the elemental rhombohedron which can be selected in the structure of γ-B. **(b.)** Single layers of the cubic closest packing (*ccp*) of icosahedra in the structures of α-B, γ-B and $B_{13}C_2$. The view is perpendicular to the layers: for α-B and $B_{13}C_2$ the direction of the view coincides with the *c*-axis of the trigonal unit cell (hexagonal settings), for γ-B it coincides with the *c´* direction (see text). **(c.)** The packing of icosahedra shown in projection along the axes of the rhombohedral cells for α-B and $B_{13}C_2$ and along the *a*-axis for γ-B. **(d.)** Fragments of the structures of α-B, γ-B and $B_{13}C_2$ showing different types of bonds in the corresponding structures (see text for details). In all structure drawings, atoms in crystallographically independent positions are marked in different colors: for α-B and $B_{13}C_2$ $B_P$ are blue, $B_E$ are green; for $B_{13}C_2$ carbon and boron atoms of the C-B-C chains are $B_C$ (violet) and C (red). For γ-B B1 are blue, B2 are green, B3 are purple, B4 are orange, B5 are violet. Different types of bonds are shown in different colors: 2e2c bonds are yellow, 1e2c bonds are magenta, 2e3c bonds are brown triangles, 3e3c bonds are black.

**Figure 2. Pressure-dependent evolution of the relative lattice parameters and the relative unit cell and $B_{12}$ icosahedron volumes for $B_{13}C_2$.** Continuous lines show the fit of the corresponding data with the third-order Birch-Murnaghan equation of state.

**Figure 3. Pressure-dependent evolution of bonds of $B_{13}C_2$.** **(a.)** Intraicosahedral bonds; **(b.)** Intericosahedral bonds. Bonds designations are according to Mondal et al.[13], corresponding notations from Dera et al.[10] (b1 through b7) are given in brackets to make it easier to compare.

**Figure 4. Relative change of interatomic distances for α-B, γ-B and $B_{13}C_2$ single crystal plotted against their length at lowest pressure as revealed by in situ single-crystal X-ray diffraction.** Circles stay for bonds in α-B, triangles for γ-B, and squares for $B_{13}C_2$. Intraicosahedral bonds are outlined by the black rectangular; red, purple and orange symbols correspond to α-B, γ-B and $B_{13}C_2$, respectively. Intericosahedral bonds and those involving B-B dumbbells and C-B-C chains are shown in green, blue, and black colors for α-B, γ-B and $B_{13}C_2$, respectively. The bonds in γ-B are shown in triangle: intraicosahedral ones (within black rectangular) are designated in purple, between icosahedra are in blue.

**Figure 5. Evolution of the ratios of the unit cell parameters *c/a* of α-B and $B_{13}C_2$ compared to that of *c´/a´* of γ-B.** For α-B the data are taken from Chuvashova et al.[25] for γ-B from Zarechnaya et al.[9]

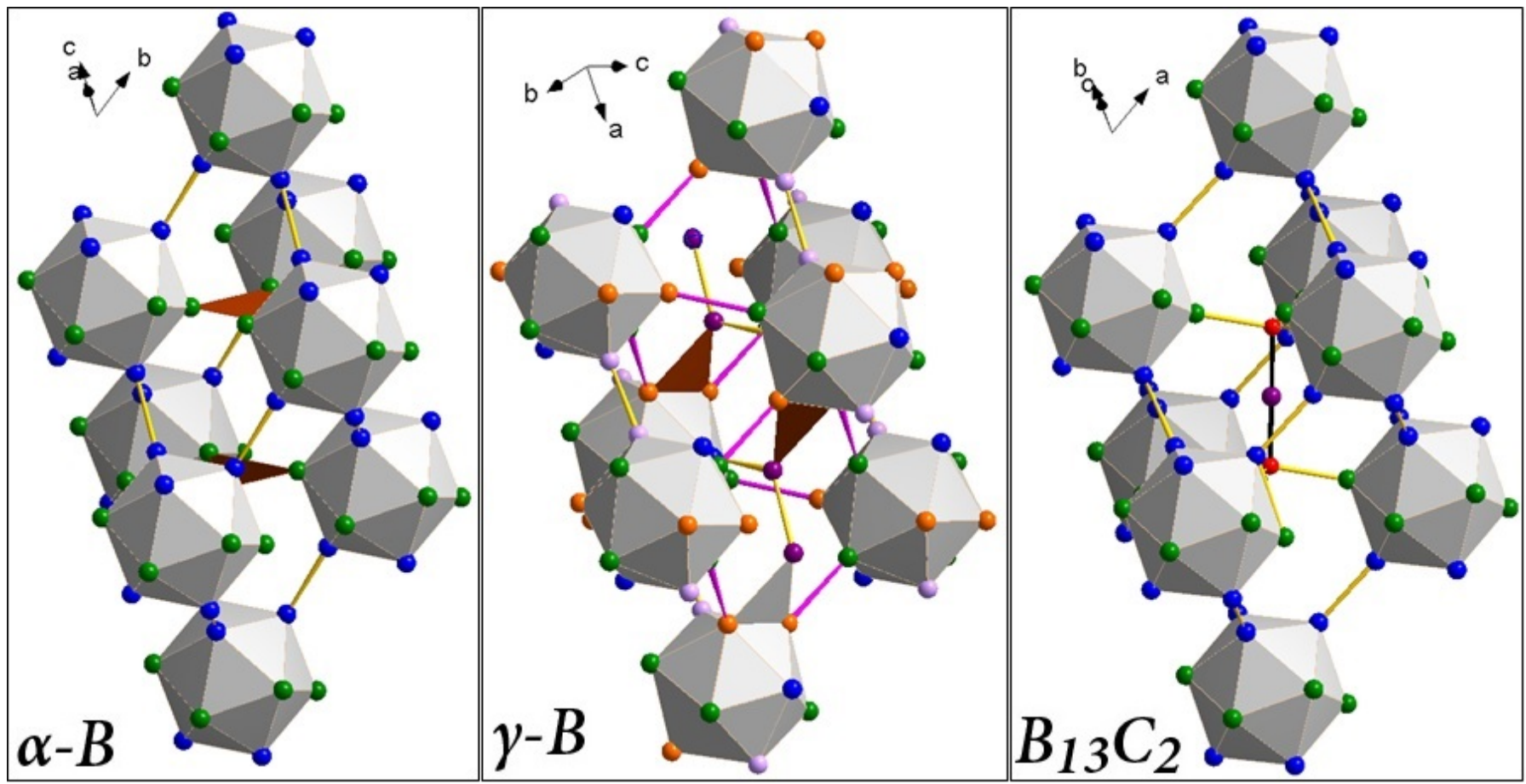

**Figure 1 a.**

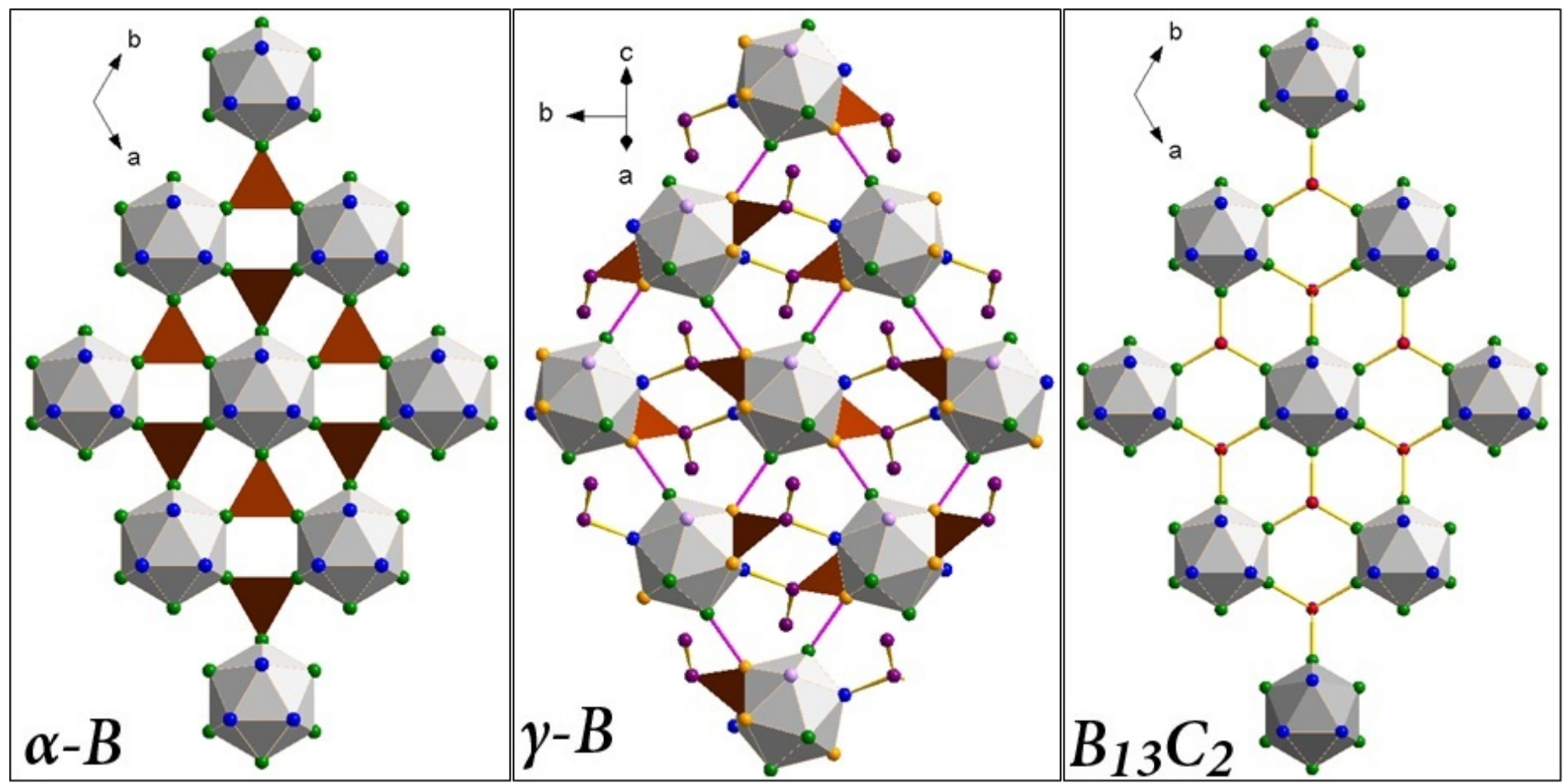

**Figure 1 b.**

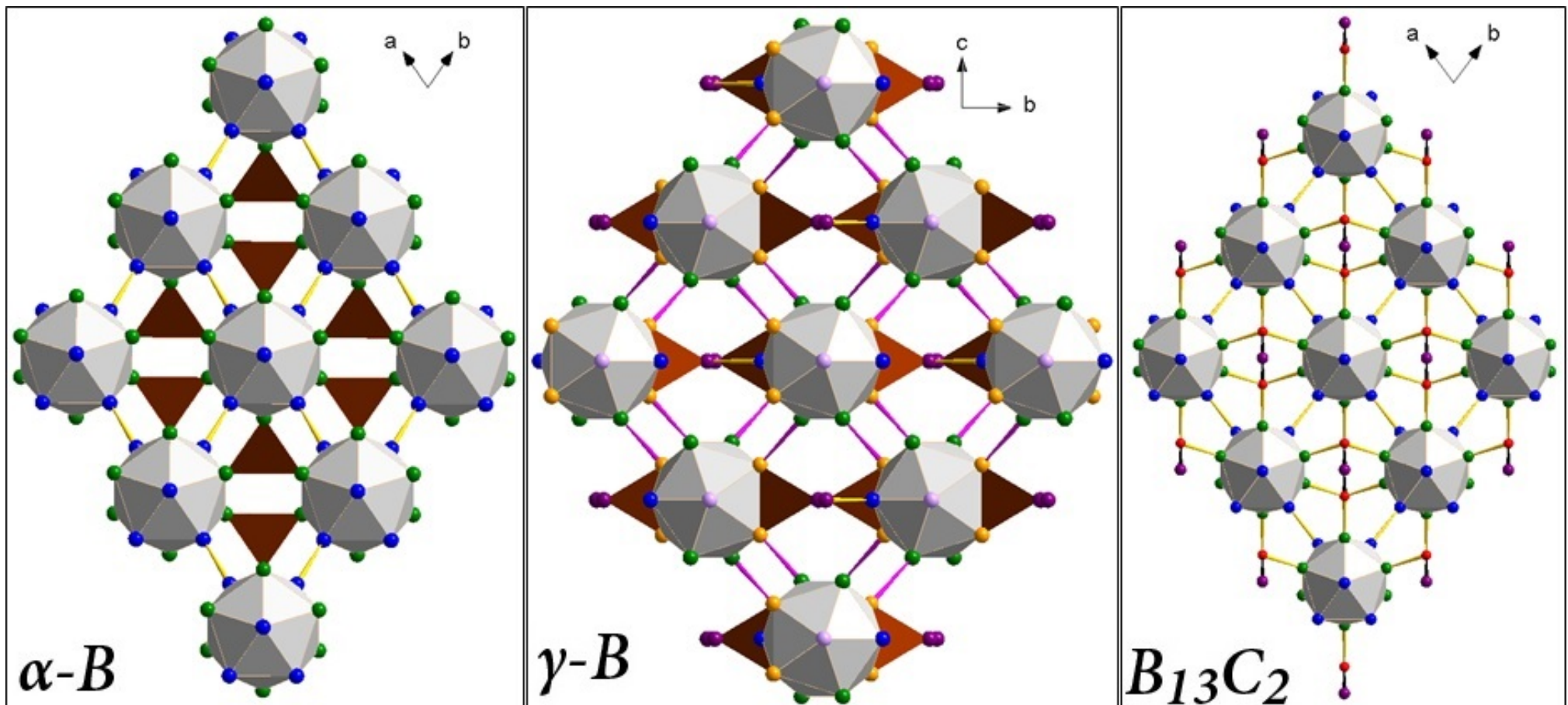

**Figure 1 c.**

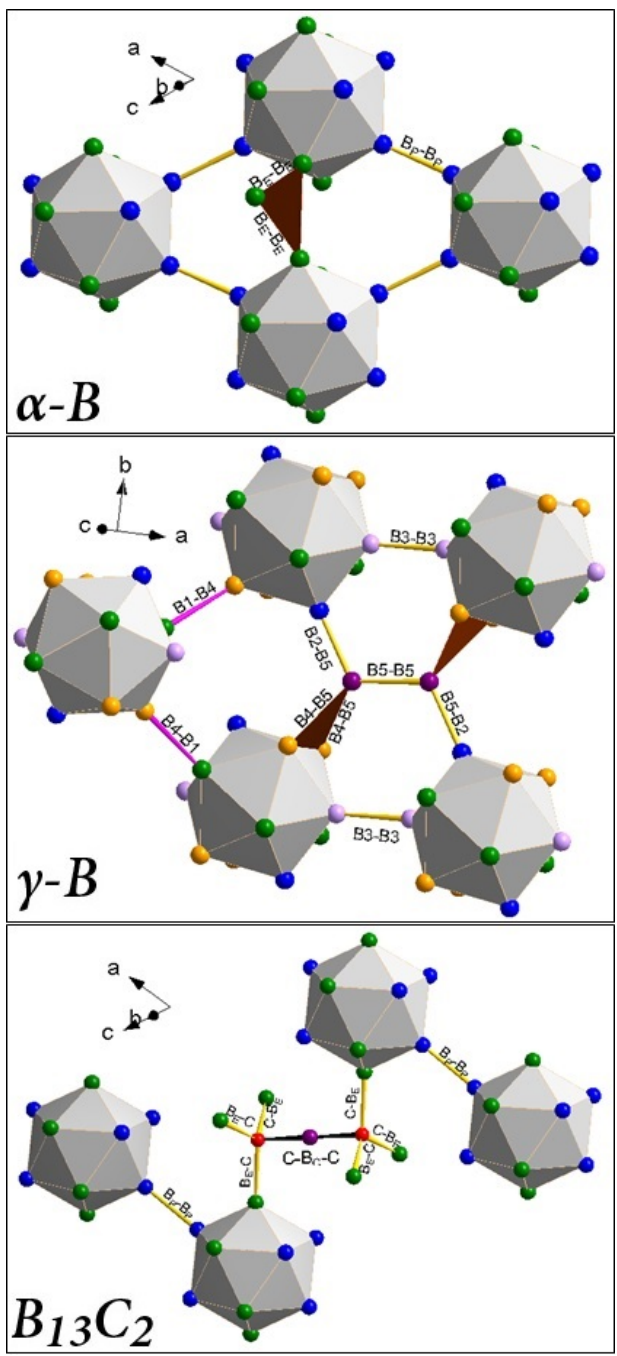

**Figure 1 d.**

**Figure 1. Crystal structures of α-B, γ-B and $B_{13}C_2$. (a.)** Rhombohedral cells of α-B and $B_{13}C_2$ compared to the elemental rhombohedron which can be selected in the structure of γ-B. **(b.)** Single layers of the cubic closest packing (*ccp*) of icosahedra in the structures of α-B, γ-B and $B_{13}C_2$. The view is perpendicular to the layers: for α-B and $B_{13}C_2$ the direction of the view coincides with the *c*-axis of the trigonal unit cell (hexagonal settings), for γ-B it coincides with the *c´* direction (see text). **(c.)** The packing of icosahedra shown in projection along the axes of the rhombohedral cells for α-B and $B_{13}C_2$ and along the *a*-axis for γ-B. **(d.)** Fragments of the structures of α-B, γ-B and $B_{13}C_2$ showing different types of bonds in the corresponding structures (see text for details). In all structure drawings, atoms in crystallographically independent positions are marked in different colors: for α-B and $B_{13}C_2$ $B_P$ are blue, $B_E$ are green; for $B_{13}C_2$ carbon and boron atoms of the C-B-C chains are $B_C$ (violet) and C (red). For γ-B B1 are blue, B2 are green, B3 are purple, B4 are orange, B5 are violet. Different types of bonds are shown in different colors: 2e2c bonds are yellow, 1e2c bonds are magenta, 2e3c bonds are brown triangles, 3e3c bonds are black.

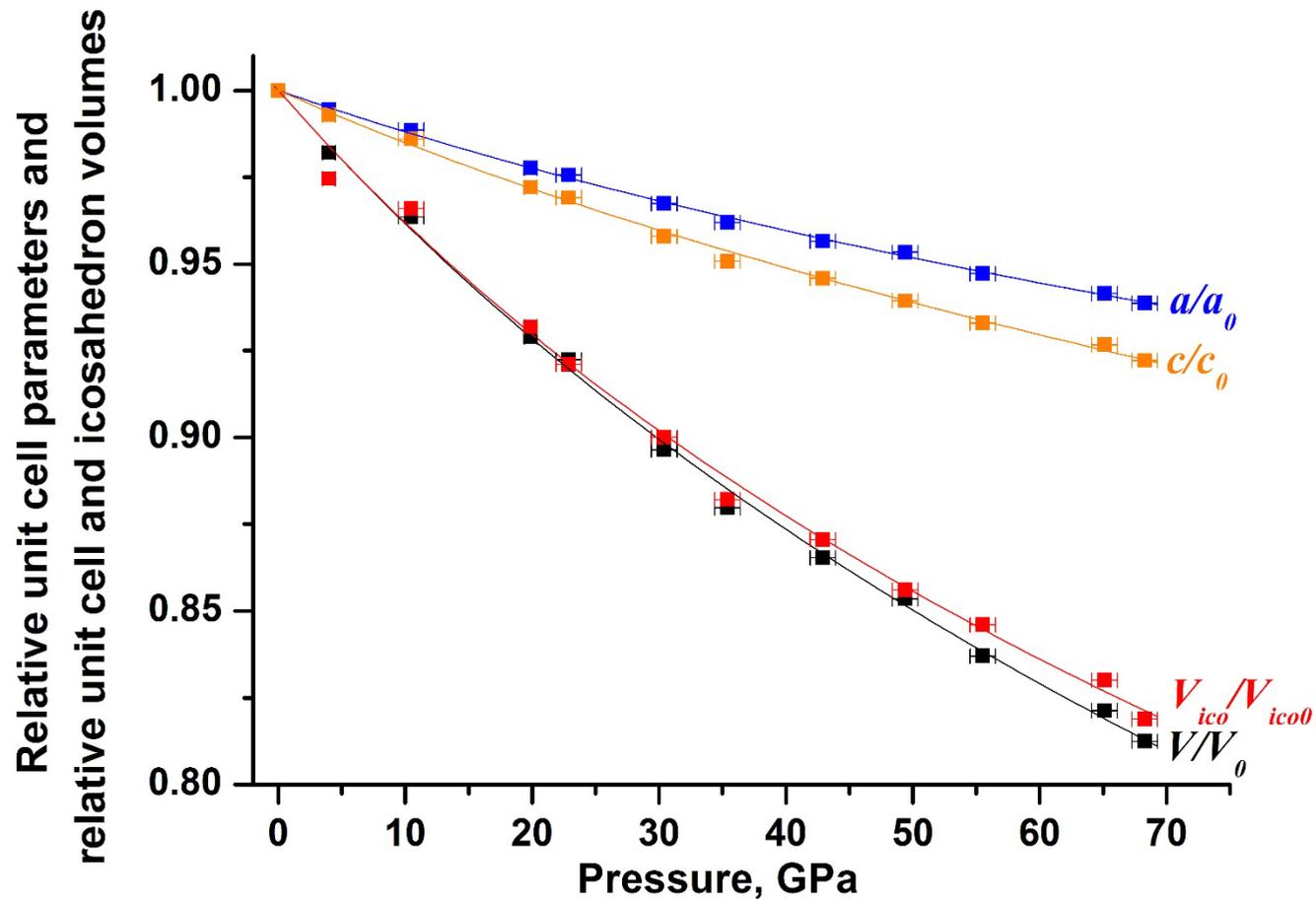

**Figure 2 Pressure-dependent evolution of the relative lattice parameters and the relative unit cell and $B_{12}$ icosahedron volumes for $B_{13}C_2$.** Continuous lines show the fit of the corresponding data with the third-order Birch-Murnaghan equation of state.

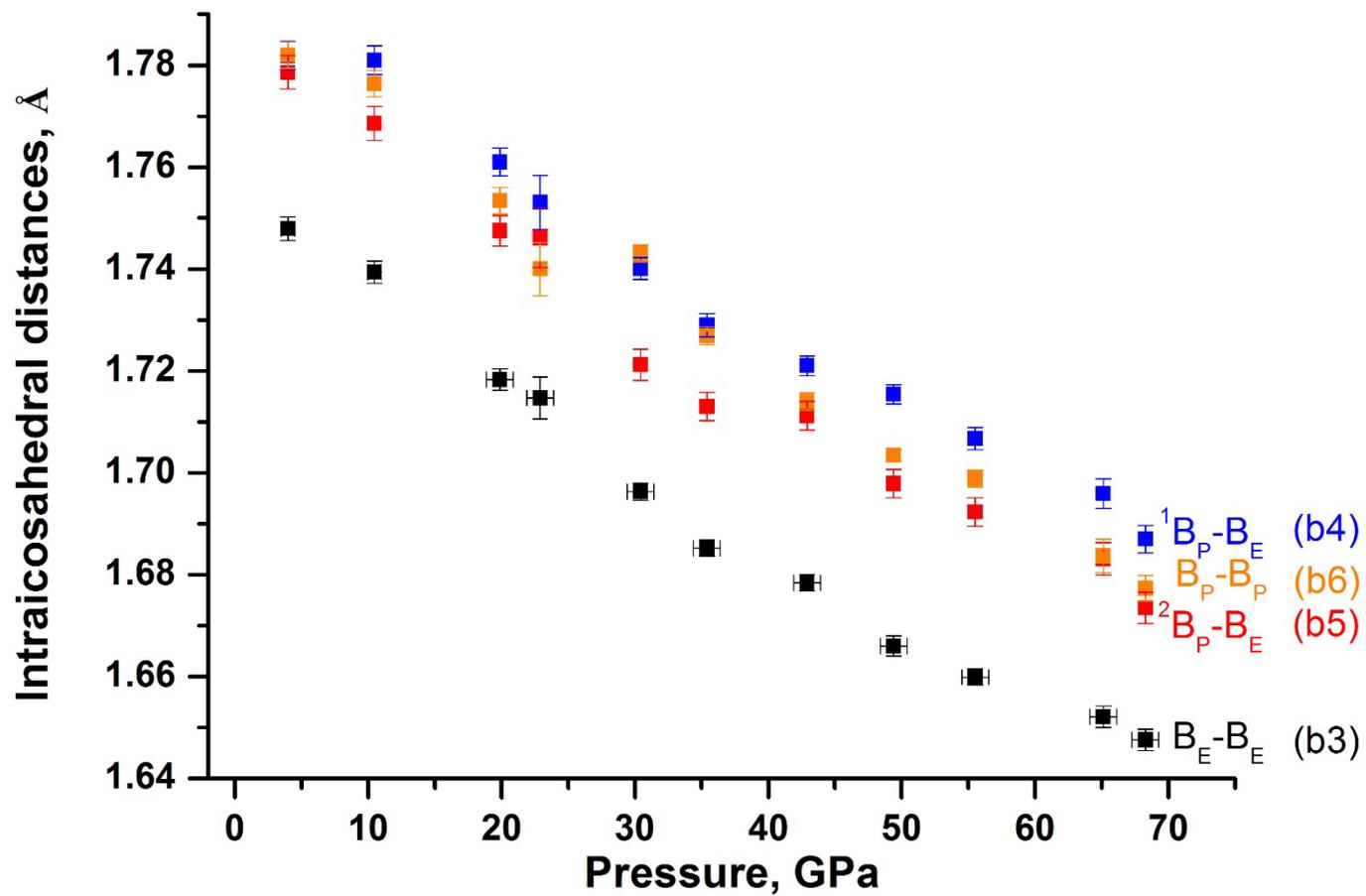

**Figure 3 a.**

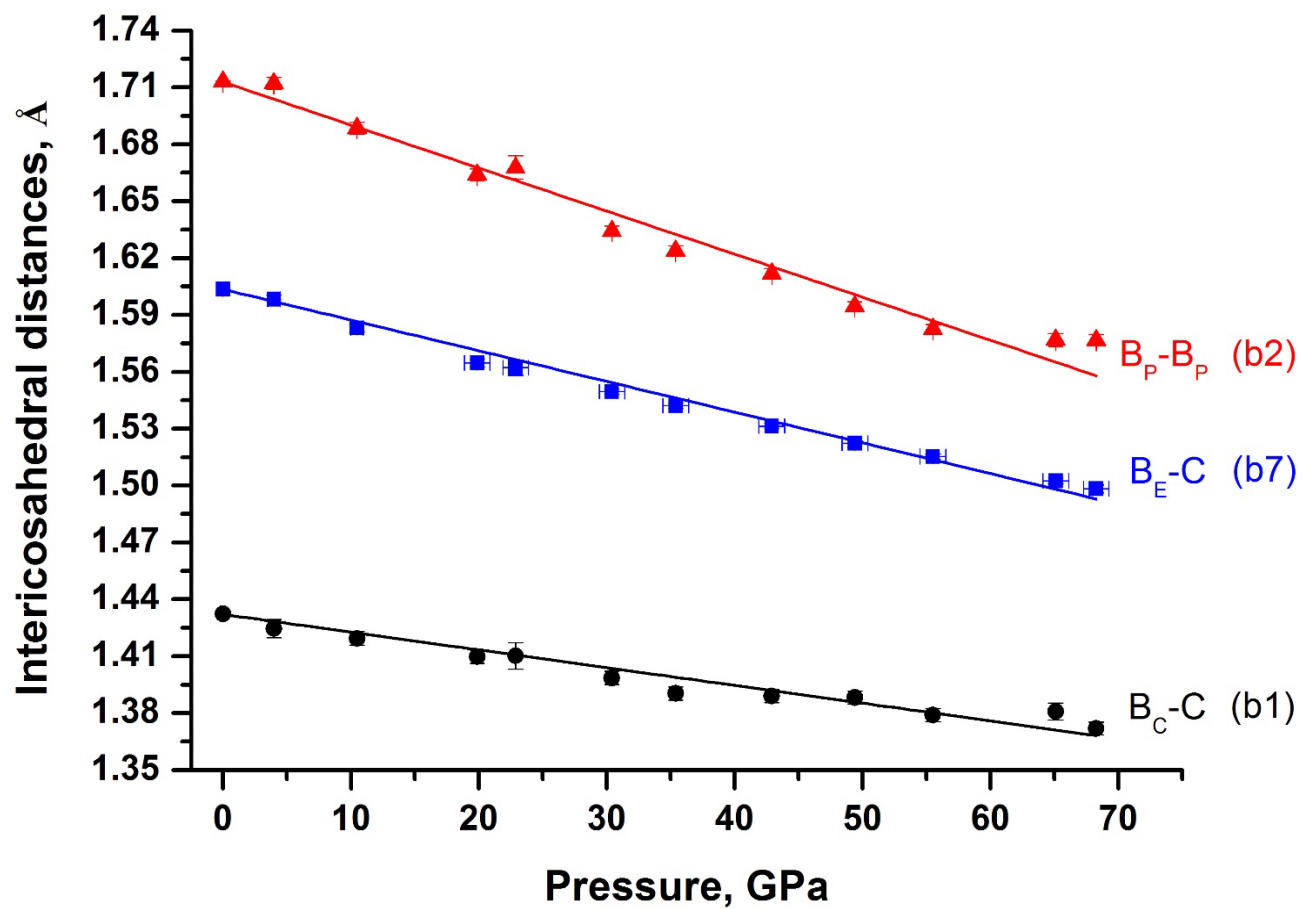

**Figure 3 b.**

**Figure 3. Pressure-dependent evolution of bonds of $B_{13}C_2$.** **(a.)** Intraicosahedral bonds; **(b.)** Intericosahedral bonds. Bonds designations are according to Mondal et al. [13]; corresponding notations from Dera et al. [10] (b1 through b7) are given in brackets to make it easier to compare.

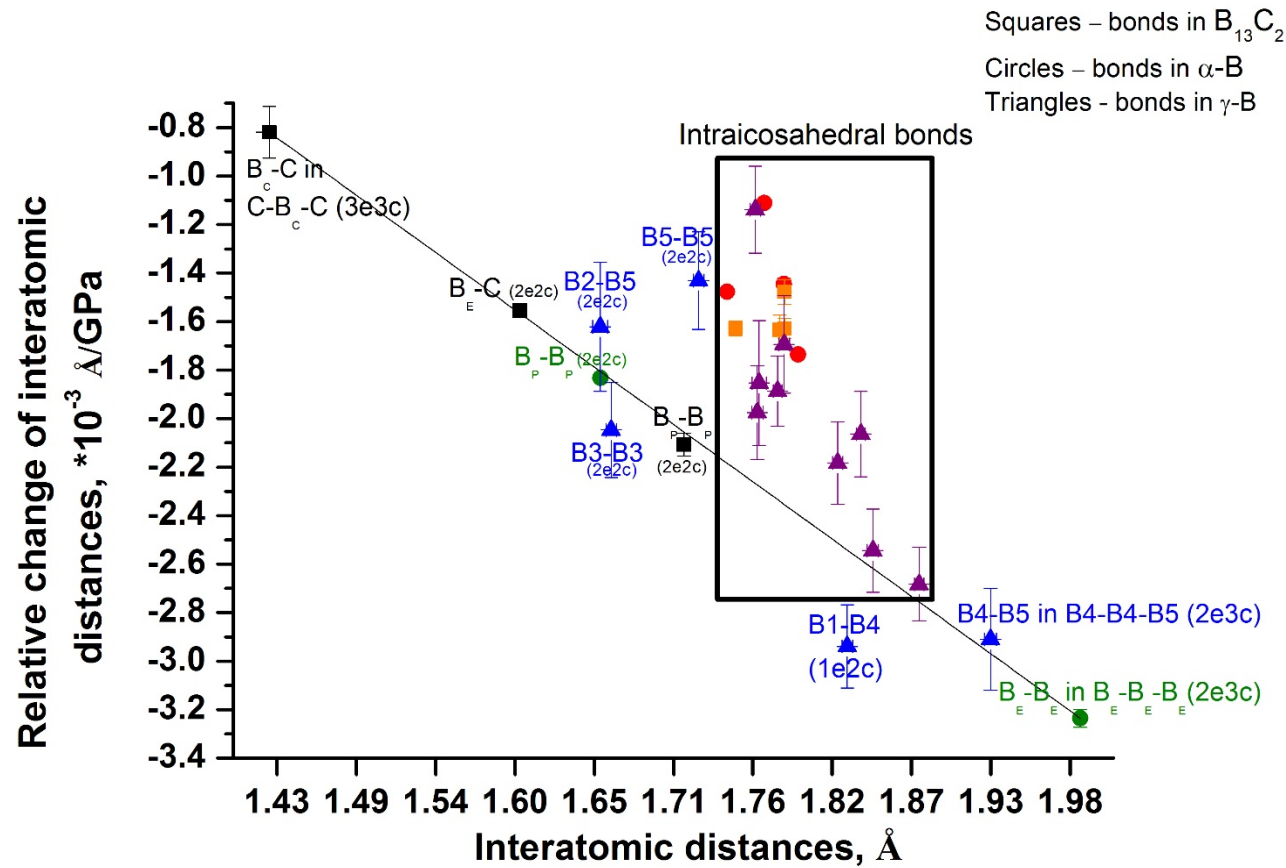

**Figure 4. Relative change of interatomic distances for α-B, γ-B and $B_{13}C_2$ single crystal plotted against their length at lowest pressure as revealed by in situ single-crystal X-ray diffraction.** Circles stay for bonds in α-B, triangles for γ-B, and squares for $B_{13}C_2$. Intraicosahedral bonds are outlined by the black rectangular; red, purple and orange symbols correspond to α-B, γ-B and $B_{13}C_2$, respectively. Intericosahedral bonds and those involving B-B dumbbells and C-B-C chains are shown in green, blue, and black colors for α-B, γ-B and $B_{13}C_2$, respectively. The bonds in γ-B are shown in triangle: intraicosahedral ones (within black rectangular) are designated in purple, between icosahedra are in blue.

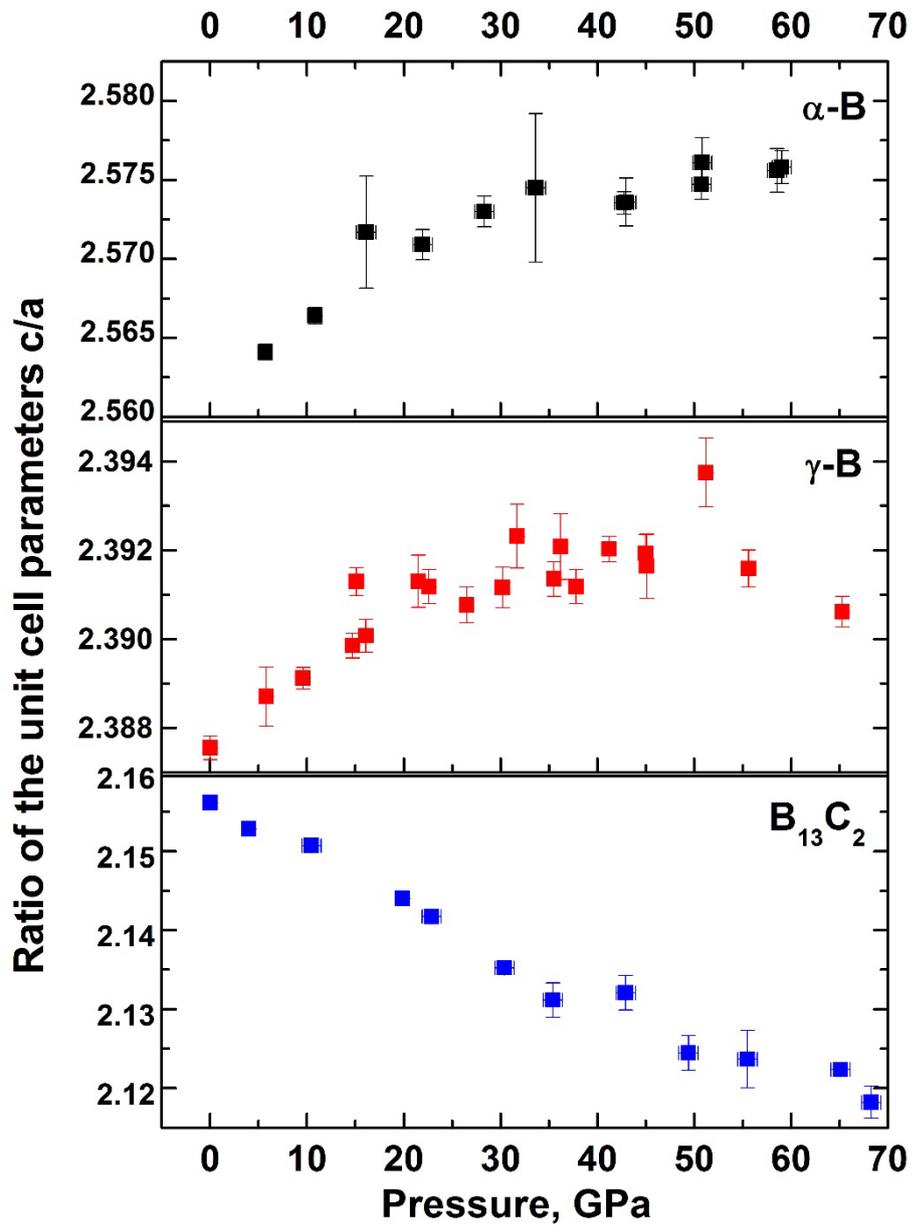

**Figure 5. Evolution of the ratios of the unit cell parameters *c/a* of α-B and $B_{13}C_2$ compared to that of *c´/a´* of γ-B.** For α-B the data are taken from Chuvashova et al.[25], for γ-B from Zarechnaya et al.[9]

**Table 1. Crystallographic details of $B_{13}C_2$ at variable pressure.** Formula: $B_{13}C_2$; Crystal System: Trigonal; Z=3; Space group: $R\bar{3}m$, crystal size $10 \times 10 \times 15 \mu m^3$ (crystal 1) and $20 \times 15 \times 10 \mu m^3$ (crystal 2, designated as "cr2").

| P, GPa | a, Å | c, Å | V, Å$^3$ | $R_{int}$* | $R_1[I>3\sigma(I)]$* | $N_m$** | $N_i$** |
|---|---|---|---|---|---|---|---|
| 0.00010(1)[13] | 5.5962(3) | 12.0661(7) | 327.25(3) | 0.0197 | 0.0227 | 505 | 440 |
| 4.0(5) | 5.5653(6) | 11.9811(11) | 321.37(6) | 0.093 | 0.0755 | 272 | 155 |
| 10.0(5) | 5.5321(9) | 11.898(2) | 315.33(9) | 0.024 | 0.0762 | 221 | 150 |
| 20(1) | 5.4708(5) | 11.7296(9) | 304.03(4) | 0.025 | 0.0577 | 362 | 149 |
| 23(1) | 5.4597(12) | 11.6932(19) | 301.86(11) | 0.024 | 0.1096 | 202 | 148 |
| 30(1) cr2 | 5.4134(8) | 11.5589(18) | 293.35(8) | 0.023 | 0.0659 | 291 | 123 |
| 35(1) cr2 | 5.383(4) | 11.472(8) | 287.9(3) | 0.030 | 0.0595 | 316 | 122 |
| 43(1) cr2 | 5.353(4) | 11.413(8) | 283.2(4) | 0.028 | 0.0617 | 268 | 117 |
| 49(1) cr2 | 5.335(4) | 11.334(8) | 279.3(3) | 0.024 | 0.0584 | 247 | 119 |
| 56(1) cr2 | 5.3007(9) | 11.257(2) | 273.91(9) | 0.024 | 0.0617 | 236 | 121 |
| 65(1) | 5.2683(10) | 11.1812(16) | 268.75(8) | 0.055 | 0.0816 | 184 | 100 |
| 68(1) cr2 | 5.2529(50) | 11.1267(10) | 265.88(4) | 0.024 | 0.0645 | 200 | 109 |

*$R_{int}$ and $R_1[I > 3\sigma(I)]$ relate to data collection and structure refinement, respectively.

**$N_m$ is the number of measured unique reflections; $N_i$ - the number of unique reflections with $I>3\sigma(I)$

**Table 2. Parameters of the equations of state of $B_{13}C_2$, $B_4C$, α-B, and γ-B obtained on the basis of single-crystal XRD in comparison with those of icosahedra.** Volume reduction of the unit cells and icosahedra on compression was calculated for the pressure range from ambient to 60 GPa.

| Material | | Ref. | $K_{300}$, GPa | K´ | $K_{ico}$, GPa | $K´_{ico}$ | Volume reduction, % | |
|---|---|---|---|---|---|---|---|---|
| | | | | | | | Unit cell | Icosahedron |
| boron carbide | $B_{13}C_2$ | *Present study* | *239(7)* | *3.2(3)* | *239(23)* | *3.8(8)* | *18.7* | *18.1* |
| | $B_4C$ | Dera et al. (2014)[10] | 243(3) | 3.6(2) | | | 18 | 13 |
| α-B | | Chuvashova et al. (2017)[25] | 224(7) | 3.0(3) | 273(12) | 4(fixed) | 18.0 | 14.5 |
| γ-B | | Zarechnaya et al. (2010)[9], P<40 GPa | 227(3) | 2.5(2) | 285(6) | 1.8(3) | 18.9 | 16.9 |

**Table 3. The $B_{12}$ icosahedron volume and bond lengths in boron carbide at variable pressure.** "cr2" designates crystal 2. $B_P$, $B_E$ and $B_C$ are designations from Mondal et al.[13]; b1 through b7 are designations from Dera et al.[10]

| P, GPa | $V_{ico}$, Å$^3$ | Bonds involving CBC | | Inter-cluster bonds | Intra-cluster bonds | | | |
|---|---|---|---|---|---|---|---|---|
| | | $B_E$-C (b7) | $B_C$-C (b1) | $B_P$-$B_P$ (b2) | $B_P$–$B_P$ (b6) | $^1B_P$–$B_E$ (b4) | $^2B_P$–$B_E$ (b5) | $B_E$–$B_E$ (b3) |
| 0.00010(1)[13] | 12.501 | 1.6037(2) | 1.4324(5) | 1.7131(4) | 1.8053(4) | 1.7997(4) | 1.7848(5) | 1.7590(3) |
| 4.0(5) | 12.1908 | 1.5980(17) | 1.4246(48) | 1.7112(32) | 1.7825(26) | 1.7826(28) | 1.7786(33) | 1.7480(23) |
| 10.0(5) | 12.0831 | 1.5828(16) | 1.4194(36) | 1.6874(31) | 1.7769(26) | 1.7817(28) | 1.7686(33) | 1.7394(22) |
| 20(1) | 11.6379 | 1.5653(21) | 1.4098(35) | 1.6649(30) | 1.7523(26) | 1.7605(27) | 1.7474(30) | 1.7177(21) |
| 23(1) | 11.5144 | 1.5627(41) | 1.4102(70) | 1.668(62) | 1.7406(52) | 1.7525(53) | 1.7469(58) | 1.7147(41) |
| 30(1) cr2 | 11.2751 | 1.5476(15) | 1.3963(35) | 1.6346(26) | 1.7420(17) | 1.7412(21) | 1.7232(31) | 1.6989(17) |
| 35(1) cr2 | 11.0331 | 1.5418(11) | 1.3904(34) | 1.6234(26) | 1.7274(17) | 1.7292(22) | 1.7130(28) | 1.6855(16) |
| 43(1) cr2 | 10.8796 | 1.5321(11) | 1.3890(34) | 1.6113(26) | 1.7146(17) | 1.7207(22) | 1.7109(28) | 1.6778(16) |
| 49(1) cr2 | 10.7212 | 1.5222(11) | 1.3871(34) | 1.5915(24) | 1.7057(17) | 1.7160(21) | 1.6990(28) | 1.6667(16) |
| 56(1) cr2 | 10.5721 | 1.5159(11) | 1.3790(34) | 1.5821(24) | 1.6994(17) | 1.7064(22) | 1.6921(28) | 1.6594(16) |
| 65(1) | 10.4294 | 1.5016(15) | 1.3775(45) | 1.5689(34) | 1.6890(33) | 1.6974(35) | 1.6867(40) | 1.6545(29) |
| 68(1) cr2 | 10.2436 | 1.498(2) | 1.3719(33) | 1.5757(30) | 1.6778(25) | 1.6876(27) | 1.6735(31) | 1.6477(21) |

**Supplementary Table S1. Fractional atomic coordinates and equivalent isotropic displacement parameters for $B_{13}C_2$ at variable pressure.** (Crystal 2 is designated as "cr2").
*Wyckoff positions are *18h* (x, $\bar{x}$, z) for both $B_P$ and $B_E$ atoms, *3a* (0,0, 0) for $B_C$ atom and *6c* (0, 0, z) for C atom.

| P, GPa | Atom* | x | z | $U_{eq}$, Å$^2$ |
|---|---|---|---|---|
| 4.0(5) | $B_P$ | 0.4401(3) | 0.0527(2) | 0.0028(10) |
| | $B_E$ | 0.5032(3) | 0.1922(2) | 0.0033(10) |
| | $B_C$ | 0 | 0 | 0.0091(19) |
| | C | 0 | 0.1189(4) | 0.0036(11) |
| 10.0(5) | $B_P$ | 0.4404(3) | 0.0522(2) | 0.0101(10) |
| | $B_E$ | 0.5038(3) | 0.1918(2) | 0.0102(10) |
| | $B_C$ | 0 | 0 | 0.0123(17) |
| | C | 0 | 0.1193(3) | 0.0097(11) |
| 20(1) | $B_P$ | 0.4401(3) | 0.05192(19) | 0.0070(7) |
| | $B_E$ | 0.5036(3) | 0.19179(18) | 0.0075(8) |
| | $B_C$ | 0 | 0 | 0.0120(15) |
| | C | 0 | 0.1202(3) | 0.0074(9) |
| 23(1) | $B_P$ | 0.4396(6) | 0.0519(4) | 0.0156(15) |
| | $B_E$ | 0.5034(6) | 0.1921(3) | 0.0156(15) |
| | $B_C$ | 0 | 0 | 0.020(3) |
| | C | 0 | 0.1206(6) | 0.0162(18) |
| 30(1) cr2 | $B_P$ | 0.4406(2) | 0.05175(19) | 0.0029(9) |
| | $B_E$ | 0.5036(2) | 0.1918(2) | 0.0032(9) |
| | $B_C$ | 0 | 0 | 0.0071(15) |
| | C | 0 | 0.1208(3) | 0.0036(9) |
| 35(1) cr2 | $B_P$ | 0.4403(2) | 0.05150(19) | 0.0076(8) |
| | $B_E$ | 0.5032(2) | 0.19180(17) | 0.0075(8) |
| | $B_C$ | 0 | 0 | 0.0144(15) |
| | C | 0 | 0.1212(3) | 0.0088(9) |
| 43(1) cr2 | $B_P$ | 0.4401(2) | 0.05114(19) | 0.0074(8) |
| | $B_E$ | 0.5032(2) | 0.19201(17) | 0.0072(8) |
| | $B_C$ | 0 | 0 | 0.0147(15) |
| | C | 0 | 0.1217(3) | 0.0079(9) |
| 49(1) cr2 | $B_P$ | 0.4401(2) | 0.05062(18) | 0.0055(8) |
| | $B_E$ | 0.5033(2) | 0.19167(18) | 0.0061(8) |
| | $B_C$ | 0 | 0 | 0.0153(15) |
| | C | 0 | 0.1226(3) | 0.0075(9) |
| 56(1) cr2 | $B_P$ | 0.4402(2) | 0.05059(18) | 0.0041(8) |
| | $B_E$ | 0.5032(2) | 0.19185(19) | 0.0050(8) |
| | $B_C$ | 0 | 0 | 0.0151(16) |
| | C | 0 | 0.1225(3) | 0.0060(9) |
| 65(1) | $B_P$ | 0.4402(4) | 0.0504(2) | 0.0044(11) |
| | $B_E$ | 0.5036(4) | 0.1921(3) | 0.0057(12) |
| | $B_C$ | 0 | 0 | 0.016(2) |
| | C | 0 | 0.1232(4) | 0.0069(12) |
| 68(1) cr2 | $B_P$ | 0.4398(3) | 0.0509(2) | 0.0026(9) |
| | $B_E$ | 0.5035(3) | 0.1920(2) | 0.0024(9) |
| | $B_C$ | 0 | 0 | 0.0153(18) |
| | C | 0 | 0.1233(3) | 0.0036(10) |

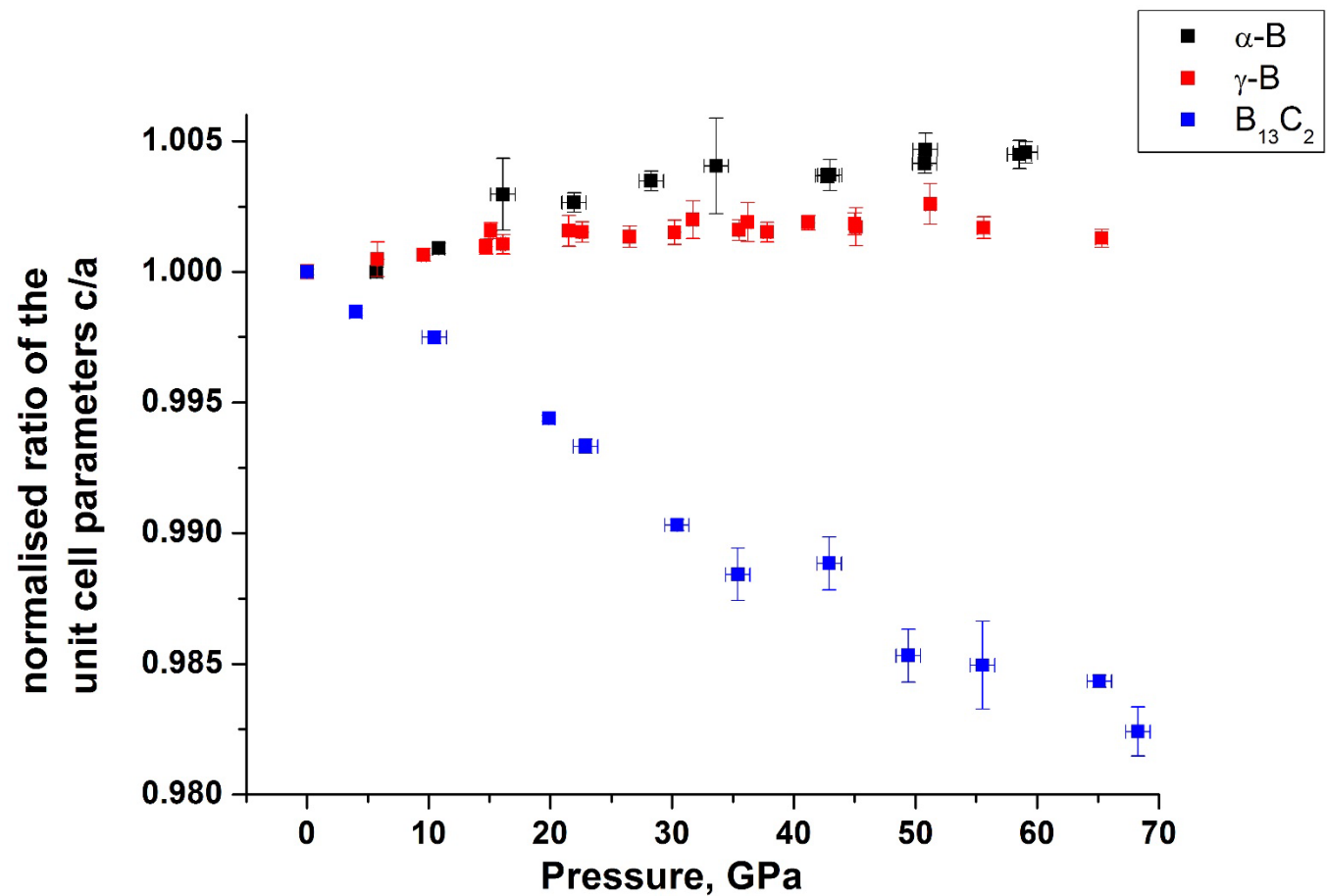

**Supplementary Figure S1. Evolution of the normalized ratios of the unit cell parameters *c/a* of α-B and B$_{13}$C$_2$ compared to that of *c´/a´* of γ-B.** For α-B the data are taken from Chuvashova et al.[25]; for γ-B from Zarechnaya et al.[9]